\documentstyle[sprocl]{article}

\input{epsfig1.sty}

\def\drop#1{}

\def\cc{\widetilde{c}}

\def\cut#1{}
\newcommand{\equ}[1]{~Eq.~(\ref{#1})}\newcommand{\no}[1]{~(\ref{#1})}

\newcommand{\sla}{\raise.15ex\hbox{$/$}\kern -.8em}

\newcommand{\half}{\frac 1 2}

\newcommand{\m}{\mu}

\newcommand{\s}{\sigma}

\newcommand{\be}[1]{\begin{equation}\label{#1}}
\newcommand{\ee}{\end{equation}}
\newcommand{\ba}[1]{\begin{eqnarray}\label{#1}}
\newcommand{\ea}{\end{eqnarray}}

\title{Pricing European Options in Realistic Markets}
\author{Martin Schaden$^\dagger$}
\address{New York University, 4 Washington Place, New York, New York 10003}
\date{\today}
\begin{document}
\maketitle
\begin{abstract}
\noindent We investigate the relation between the fair price for
European-style vanilla options and the distribution of short-term
returns on the underlying asset ignoring transaction and other
costs.  We compute the risk-neutral probability density
conditional on the total variance of the asset's returns when the
option expires. If the asset's future price has finite
expectation, the option's fair value satisfies a parabolic partial
differential equation of the Black-Scholes type in which the
variance of the asset's returns rather than a trading time is the
evolution parameter. By immunizing the portfolio against
large-scale price fluctuations of the asset, the valuation of
options is extended to the realistic case\cite{St99} of assets
whose short-term returns have finite variance but very large, or
even infinite, higher moments. A dynamic Delta-hedged portfolio
that is statically insured against exceptionally large
fluctuations includes at least two different options on the asset.
The fair value of an option in this case is determined by a
universal drift function that is common to all options on the
asset. This drift is interpreted as the premium for an investment
exposed to risk due to exceptionally large variations of the
asset's price. It affects the option valuation like an effective
cost-of-carry for the underlying in the Black-Scholes world would.
The derived pricing formula for options in realistic markets is
arbitrage free by construction. A simple model with constant drift
qualitatively reproduces the often observed volatility -skew and
-term structure.

\end{abstract}
\footnotetext{$^\dagger$ Email address: m.schaden@att.net}

\section{Introduction}
An important result of modern finance is that the fair
(no-arbitrage) price $V$ for a European-style option is the
expected present value (PV) of its future payoff,
\be{fta}
V={\bf E}^Q[PV({\rm payoff})] \ .
\ee
The expectation in\equ{fta} is with respect to a risk-neutral
(martingale) measure $Q$ on the space of price-paths. The
fundamental theorem of asset pricing\cite{Sc92} ensures the
existence of the risk-neutral measure $Q$ in the absence of
arbitrage opportunities, but does not explicitly relate it to the
process for the underlying. We shall see that in some cases of
interest, the validity of\equ{fta} is restricted to options with a
bounded payoff.

Instead of directly computing the expectation in\equ{fta} for
European-style options, we will consider the conditional
expectation
\be{condexp}
{\bf E}^Q[PV({\rm payoff})|v_f]\ ,
\ee
with respect to volatility paths with the same {\it total} final
variance $v_f$ of the asset's returns. \equ{fta} is recovered by
taking the expectation over $v_f$. For asset prices that follow a
diffusion process, the conditional expectation of\equ{condexp}
turns out to be unique, but we will see that this generally is not
true for realistic processes.

The analysis of the fair value of European-style call and put
options by Black and Scholes\cite{BS73} was based on a stochastic
model in which the returns of the asset follow a random walk. The
fair value $C_{BS}$ of an European-style call in this model
depends only on the asset's spot price $S_0$, the volatility $\s$
and risk-free rate $r$ (both assumed constant), time to exercise
$T$ and the option's strike $K$. Dimensional analysis requires
that
\be{BScall}
C_{BS}(S_0,\s,r,T;K)=S_0 c_{BS}(rT, \s^2 T;K/S_0).
\ee
Note that the valuation\no{BScall} of a European-style call
depends only on the final variance $v_f=\s^2 T$, and the
integrated discount factor $rT$, rather than separately on the
volatility $\s$, risk-free rate $r$ and time to expiration $T$.
Assuming that the (mean) risk-free rate is known,\equ{BScall} can
be inverted to give the (implied) volatility $\s_{BS}^{\rm
implied} $ with which the Black-Scholes model would reproduce the
observed spot price $C$ of a call with time to expiration $T$ and
strike $K$,
\be{implied}
\s_{BS}^{\rm implied}=\tilde\s(C/S_0,rT,K/S_0)/\sqrt{T}\ ,
\ee
where $\tilde\s$ is the dimensionless overall standard deviation
of the distribution of returns. Limitations of the Black-Scholes
option pricing formula are expressed by the fact that the implied
volatility of European-type options generally is found to depend
on the strike $K$ and time to expiration $T$. The volatility
implied by calls that are in-the-money very often is higher than
that implied by out-of-the-money calls. A graph of the implied
volatility against the call's strike therefore tends to "smile"
(somewhat crookedly) rather than frown. The effect is referred to
as the volatility -smile or -skew. The dependence of the implied
volatility on $T$ is known as the volatility's term structure.

The observed volatility skew has been traced to a number of
causes. All of them are related to a higher probability for
exceptionally large fluctuations in the returns than the random
walk model admits. Stochastic models that simulate this effect
have been considered\cite{Bo86}, but a quantitative explanation of
the empirically observed fluctuations has only recently been
proposed\cite{Sc02b}. Since the observed large-scale fluctuations
in returns are not quantitatively reproduced by a simple
stochastic model it perhaps is of some interest to (re)examine the
problem of option pricing assuming as little as possible. The fair
price of an option in fact does not depend on many details of the
process for the underlying asset. Only the short-term transition
probability is relevant for a fully dynamic hedging strategy, but
this strategy is quite different for the following three kinds of
assets:
\begin{itemize}
  \item[I] The asset's expected future price is well-defined by
  the short-term transition probability of its returns.
   \item[II] The variance of the short-term returns on the asset is
   finite, but the asset's expected future price diverges.
  \item[III] The variance of the short-term returns on the asset diverges.
\end{itemize}

A normal distribution of the short-term returns is an example in
the first class, but not the only one. Any distribution of returns
that falls off sufficiently rapidly and in particular any
distribution with compact support belongs to this class.
No-arbitrage arguments uniquely price European-style options on
assets in this class if the final variance of the asset's returns
is known. It is possible to construct a dynamic portfolio with
just one kind of option (in addition to the asset) that is without
appreciable risk.

The other two classes sub-divide the category of assets with
sub-exponential short-term return distributions.
Equities\cite{St99}, indices\cite{Go99} and commodities\cite{Ma02}
historically fall in the second class of assets and this case will
therefore concern us most. It turns out that one still can
construct a dynamic portfolio that is without appreciable risk,
but the portfolio in this case includes at least two different
options on the underlying asset. A portfolio with just one option
(and the asset) cannot be insured against exceptionally large
price fluctuations of the asset and is therefore not without risk.
Although the risk-neutral conditional expectation of\equ{condexp}
exists for options with bounded payoffs, it no longer is uniquely
related to the process for the underlying.

Very little can be said about the third possibility, the Paretian
case. The construction of a risk-free dynamic portfolio from
options on the underlying is no longer possible. Indeed, the
notion that the variance of the returns is a measure of risk has
to be reexamined and\equ{fta} may not be very meaningful. Since
the variance of returns for assets on which vanilla options can be
drawn apparently is finite\cite{St99,Go99,Ma02}, the Paretian case
will not be further investigated here.

We proceed as follows. Using the variance of the asset's returns
as the evolution parameter, the Black-Scholes analysis is extended
to European-style options on any class~I asset in the next
section. In section~3 we extend the analysis to include options on
assets that belong to class~II (the realistic case). Section~4
summarizes and discusses some aspects of the results.

\section{A Variation on the Black-Scholes Analysis of Option Prices}
It is useful to slightly generalize the Black-Scholes analysis to
the case where the volatility of the underlying can be an
arbitrary function of time.

We will use the variance $v$ of the asset's returns rather than a
(continuous) trading time to parameterize the evolution of an
option's fair value. The variance is a monotonically increasing
quantity; a trading- or calender- time $t$, can be viewed as {\it
defining} an instantaneous volatility $\s(t)$:
\be{volat}
\s^2(t):=\partial v/\partial t\ge 0 .
\ee
On any given volatility path $\{\s(t);0\leq t\leq T\}$ there is a
one-to-one correspondence between the variance $v$ and the "time"
$t$.
\be{vt}
v(t)=\int_0^t d\xi \s^2(\xi).
\ee
[The origin of the time-scale here is chosen to coincide with the
moment of vanishing uncertainty in the asset's price.] \equ{vt}
enables one to formally consider the evolution in "time" as an
evolution in the variance of the underlying's returns (if the
volatility is finite).

To compensate for the time value of money, all prices will be
stated as multiples of the price of an actively traded risk-free
bond that matures when the European-style option expires. The spot
price of the bond is $S_B(t)$ and its nominal value $N_B=S_B({\rm
maturity})$. Since the transition probability is for the returns
rather than for the price of the underlying, it is convenient to
convert to the dimensionless variables,
\be{logp}
x(t):=\ln[S(t)/S_B(t)]\ ,\ k:=\ln[K/N_B]\ .
\ee
Changes in the log-price $x$ give the return on the underlying
relative to the return on the bond and $k$ is the strike value of
$x$. We assume that the fair price $C$ of a European-style call
option at any moment depends only on the time to expiration, the
strike price and the spot prices for the underlying asset and the
bond. The fair call price in multiples of $S_B$ and expressed in
the above dimensionless quantities is denoted by,
\be{defc}
c_k(x,v):=C(S(t),S_B(t),t;K)/S_B(t)\ .
\ee

\subsection{Generic Properties of the Short-Term Transition Probability}
The dynamic hedging strategy of Black and Scholes that assigns a
fair value to a European-style call depends on the existence of a
very simple portfolio that is without appreciable risk for a
sufficiently short period of time.

Let the current log-price of the stock be $x$ and the probability
that the stock will have an excess return between $y-x$ and
$y+dy-x$ a short time from now be described by the transition
probability density,
\be{pdf}
p_h(y|x,\dots)=p_h(y|x,v) .
\ee
A small variance $h$ of the transition probability corresponds to
a short time interval. The ellipses denote all additional
quantities on which the transition probability may depend, such as
market- and economic- indicators, the weather and political
environment, etc. In effect, the transition probability for the
returns depends on the current time $t$, respectively on the
variance $v$.  One fortunately does not require detailed knowledge
of $p_h(y|x,v)$ to value an option on the asset.

The transition probability\equ{pdf} furthermore depends only on
the excess return $y-x$ rather than on $x$ and $y$ individually.
This financially plausible proposition can be cast in the form of
a denominational argument: the probability for a certain change of
the asset's price should not depend on whether a single bond with
value $S_B(t)$ or a package of two, three, or for that matter
$6.378$ bonds is used as price reference. With the
definition\no{logp}, this freedom in the reference denomination
implies that the transition probability is invariant under
(global) translations $z$ of all log-prices,
\be{translinv}
p_h(y-z|x-z,v)=p_h(y|x,v)=p_h(y-x|0,v)\ ,\ \ \forall z\ ,
\ee
where the latter expression is obtained by setting $z=x$. It is
important that neither the variance $h$ of the short-term returns
nor the variance $v$ of the overall returns are affected by this
translation. Because the short-term transition probability density
$p_h$ depends only on the difference $y-x$, the variance of the
overall returns is additive: the current variance of the returns
$v$ increases to $v+h$ after the short time interval we are
considering.

Since $h\rightarrow 0_+$ as the time interval is shortened, the
transition probability density has to approach Dirac's
distribution in this limit,
\be{lim0}
\lim_{h\rightarrow 0_+} p_h(y|x,v)=\delta(y-x).
\ee
Due to\equ{lim0} the expected future log-price of the stock,
\be{mean}
\bar y(h;v)=x+\m(h;v):=\int_{-\infty}^\infty dy\, y p_h(y|x,v)\ ,
\ee
approaches the current log-price $x$ and $\m(h;v)$ must become
arbitrarily small for $h\rightarrow 0_+$.  $\m(h;v)$ is the
expected excess return on the asset at a given point in time when
the variance of the asset's returns increases by $h$. It may
appear financially reasonable to assume that $\m(h;v)$ for $h\sim
0$ has the expansion $\m(h;v)=a(v) h+b(v) h^2+\dots$. However,
ignoring transaction costs, very short-term stock investments
could have a higher expected rate of return than long-term ones.
To avoid the financially unstable situation that the return on
very short-term investments becomes absolutely certain, it is
sufficient to require that,
\be{meangrowth}
\lim_{h\rightarrow 0_+}\frac{\m(h;v)}{\sqrt{h}}=0\ .
\ee
The mean return in other words should not outstrip the width of
the distribution of short-term returns.

The second moment of the transition probability, by definition, is
given by its variance $h$ and $\bar y(h;v)$
\be{m2}
{\bf E}^{p_h}[y^2]=h+{\bar y(h;v)}^2:=\int_{-\infty}^\infty dy\,
y^2 p_h(y|x,v)\ .
\ee
Somewhat surprisingly perhaps, one does not require detailed
knowledge of the higher moments of the distribution of short-term
returns.

\subsection{The Black-Scholes Valuation of
a European-Style Call} Emulating the analysis of Black and
Scholes\cite{BS73}, the fair value of a European-style call is
found by constructing a portfolio that is without appreciable risk
for sufficiently small $h$. Consider a portfolio $P$ of one
European-style call with strike $K$ and $-\Delta$ of the
underlying\footnote{We assume that the asset can be sold short and
ignore transaction fees, dividends and other costs-of-carry as
well as bid-ask spreads.}. When the portfolio is set up at a
log-price $x$ for the asset, the value $V_P$ of this position is,
\be{initv}
V_P(x,v)=c_k(x,v)-\Delta(x,v) e^x
\ee
bonds at $S_B(v)$. If the hedge ratio $\Delta(x,v)$ is not
changed, the value of this portfolio (in bonds) when the variance
of the asset's returns has increased by $h$ becomes,
\be{finv}
V_P(y,v+h)=c_k(y,v+h)-\Delta(x,v) e^y ,
\ee
if the stock's excess return over this period is $y-x$. To avoid
arbitrage, the value of this position when the hedge is set up
should be its expected future value discounted by a factor that
accounts for the risk of investing in portfolio $P$. One thus
quite generally comes to the conclusion that,
\be{pricing}
V_P(x,v)=e^{-R_P(h,v)}\int_{-\infty}^{\infty} dy\,
p_h(y|x,v)V_P(y,v+h)\ .
\ee
The discount factor $e^{-R_P(h,v)}$ is compensation for the excess
risk associated with holding the portfolio $P$ rather than the
(risk-free) bonds [by pricing relative to the bond, we already
took the time value of money into account]. $R_P(h,v)$ depends not
only on the perceived risk of the portfolio, but also reflects the
valuation of this risk by investors. The value of a certain risk
generally depends on the circumstances, that is on the time $t$,
respectively on the overall variance $v$. $R_P(0,v)=0$ in order
for\equ{pricing} to be consistent. The absence of arbitrage
opportunities requires that $R_P(h,v)\geq 0$ for all $h$. If the
portfolio is without appreciable risk over the interval $h$,
$R_P(h,v)=0$ and\equ{pricing} becomes the martingale hypothesis.
Note that the general form of \equ{pricing} is valid for finite
$h$ and could be the starting point for valuing hedge slippage.

Since the transition probability $p_h(y|x,v)$ is strongly peaked
near $y\sim x$ for $h\rightarrow 0_+$, one is led to expand the
portfolio's future value\no{finv} about $y=x$ and $h=0$. The first
few terms of this expansion are,
\ba{expansion}
V_P(y,v+h)&=&V_P(x,v)+ (y-x)[c_k^\prime(x,v)-\Delta(x,v) e^x]+
h{\dot c}_k(x,v)\cr &&\hskip-2em
+\half(y-x)^2[c_k^{\prime\prime}(x,v)-\Delta(x,v) e^x]+
O(h(y-x),(y-x)^3,h^2)\ ,\cr &&
\ea
where the shorthand notation,
\be{defdotprime}
\dot \phi(x,v):=\frac{\partial}{\partial v} \phi(x,v)\qquad {\rm
and} \qquad \phi^\prime(x,v):=\frac{\partial}{\partial x}
\phi(x,v)\ ,
\ee
denotes partial derivatives of a function with respect to $v$ and
$x$.

The term proportional to $y-x$ in\equ{expansion} vanishes for the
particular hedge ratio\footnote{To verify that\equ{riskless} is
precisely the hedge of Black and Scholes, note that with
definition\no{logp}, $e^{-x}\frac{\partial}{\partial
x}=S_B\frac{\partial}{\partial S}$}
\be{riskless}
\Delta(x,v)\rightarrow \underline{\Delta}(x,v)=e^{-x}
c_k^\prime(x,v)\ ,
\ee
and the corresponding portfolio will be denoted by
$\underline{P}$. Its value $V_{\underline{P}}$ for $y\sim x$ and
$h\sim 0$ has the simplified expansion,
\ba{expriskless}
V_{\underline{P}}(y,v+h)&=&V_{\underline{P}}(x,v)+ h{\dot
c}_k(x,v)+\frac{(y-x)^2}{2}[c_k^{\prime\prime}(x,v)-c^\prime(x,v)]\nonumber\\
&&\qquad + O(h(y-x),(y-x)^3,h^2)\ .
\ea
We show below that the portfolio $\underline{P}$ is without
appreciable risk for sufficiently small $h$ if the short-term
return distribution of the asset belongs to class~I. It is
important that this hedge depends only on the (observed) log-price
$x$ at the time it is entered into.

Let us for the moment assume that the contribution to the integral
in \equ{pricing} from higher order terms in the
expansion\equ{expansion} becomes negligible for $h\rightarrow 0_+$
(see Appendix~A). In this case the hedge\no{riskless} allows us to
evaluate the RHS of\equ{pricing} for sufficiently small $h$ as,
\ba{BSvar}
V_{\underline{P}}(x,v)&=& e^{-R_{\underline{P}} (h,v)}
\int_{-\infty}^{\infty} dy\, p_h(y|x,v)V_{\underline{P}}(y,v+h)\cr
&\sim& V_{\underline{P}}(x,v)-
R_{\underline{P}}(h,v)V_{\underline{P}}(x,v)\cr && + h{\dot
c_k}(x,v)+\frac{h+\m^2(h;v)}{2}[c_k^{\prime\prime}(x,v)-c_k^\prime(x,v)]\
,\cr &&
\ea
where we have used that by Eqs.\no{mean} and\no{m2},
\be{expxy}
h+\m^2(h;v)=\int_{-\infty}^{\infty}dy\, (y-x)^2 p_h(y|x,v)\ .
\ee
Taking the limit $h\rightarrow 0_+$ of\equ{BSvar} and
using\equ{meangrowth}, the fair option price $c_k(x,v)$ is seen to
satisfy the partial differential equation,
\be{diffBS}
{\dot c}_k(x,v)+\half[c_k^{\prime\prime}(x,v)-c_k^\prime(x,v)]=
r_{\underline{P}}(v)[c_k(x,v)-c_k^\prime(x,v)]\ ,
\ee
with a mean excess portfolio return per unit of variance of,
\be{rtl}
r_{\underline{P}}(v)=\lim_{h\rightarrow 0_+}
 h^{-1} R_{\underline{P}}(h;v)\ .
\ee

Note that $c_{-\infty}(x,v)=e^x$, is a particular solution
to\equ{diffBS}, because a call with strike $K=0$ has the same
intrinsic value as the underlying. It should be emphasized that
the mean return $\m(h;v)$ of the underlying asset does not
enter\equ{diffBS} as long as it satisfies \equ{meangrowth}.

Using the definitions\no{logp} and\no{defc} and assuming that the
volatility is a known function of the trading time $t$,
\equ{diffBS} assumes a more familiar form when the evolution is
parameterized by $t$,
\be{BS}
\left[\frac{\partial}{\partial t} + \tilde r_{\underline{P}}(t)
S\frac{\partial}{\partial
S}+\frac{\s^2(t)}{2}S^2\frac{\partial^2}{\partial
S^2}\right]C(S,t;K)=\tilde r_{\underline{P}}(t) C(S,t;K)\ .
\ee
The portfolio's instantaneous overall return rate $\tilde
r_{\underline{P}}(t)$ in\equ{BS} consists of two parts: the
(risk-free) return rate of the bond $r(t)=S_B^{-1} d S_B(t)/dt$
and the risk-premium of the portfolio,
\be{tilder}
\tilde r_{\underline{P}}(t)=r(t)+\s^2(t) r_{\underline{P}}(v(t))\
.
\ee
\equ{BS} is the partial differential equation of Black and
Scholes\cite{BS73} for the valuation of options with an in general
time-dependent volatility and an option-dependent discount rate
$\tilde r_{\underline{P}}(t)$. However, \equ{diffBS} is not
merely\equ{BS} rewritten in terms of other variables:\equ{diffBS}
remains valid even if the volatility is stochastic or an unknown
function of the trading time. \equ{diffBS} can still be integrated
in this case and shows that the fair value of a European-style
option depends only on the overall variance of the asset's returns
when it expires.

We have yet to show that the portfolio $\underline{P}$ with the
hedge ratio\no{riskless} is (at least formally) without
appreciable risk and that $r_{\underline{P}}(v)$ therefore
vanishes in the absence of arbitrage opportunities. $\tilde
r_{\underline{P}}(t)=r(t)$ in\equ{BS} then does not depend on the
option and becomes the risk-free rate of the bond.

\subsection{The Risk of Holding the Dynamically Hedged Portfolio $\underline{P}$}
A portfolio is without appreciable short-term risk compared to an
investment in the asset alone, if the variance of the portfolio's
return decreases faster than the variance of the asset's return,
which is $h$. One thus has to show that
\be{risklessdef}
\lim_{h\rightarrow 0_+} h^{-1}{\bf
Var}[V_{\underline{P}}(y,v+h)]=\frac{d}{dv}{\bf Var}[
V_{\underline{P}}(y,v)]=0
\ee

We continue to assume (see Appendix~A for details) that the
transition probability is sufficiently sharply peaked about $y\sim
x$ and again expand,
\ba{expv2}
V_{\underline{P}}^2(y,v\!+\!h)&=& V_{\underline{P}}^2(x,v)+
2V_{\underline{P}}(x,v)\left\{h{\dot
c}_k(x,v)+\frac{(y\!-\!x)^2}{2}[c_k^{\prime\prime}(x,v)-c_k^\prime(x,v)]\right\}\cr
&&\qquad + O(h(y\!-\!x),(y\!-\!x)^3,h^2)\ .
\ea
The expectation of $V_{\underline{P}}^2(y,v+h)$ to order $h$ then
is,
\ba{expV2}
{\bf E}^{p_h}[V^2_{\underline{P}}]:&=&\int_{-\infty}^\infty dy\,
p_h(y|x,v)V_{\underline{P}}^2(y,v+h)\cr &=&
V_{\underline{P}}^2(x,v)+2V_{\underline{P}}(x,v)\left\{h{\dot
c}_k(x,v)+\frac{h+\m^2(h)}{2}[c_k^{\prime\prime}(x,v)-c^\prime(x,v)]\right\}\cr
&&\hskip3in +O(h^{3/2})\cr
 &=&{\bf E}^{p_h}[V_{\underline{P}}]^2+O(h^{3/2}) \ .
\ea
The variance of the returns on portfolio $\underline{P}$ therefore
is of order $h^{3/2}$ and the risk of entering into this
investment compared to an investment in the underlying can
theoretically be made as small as one wishes by rebalancing the
portfolio often enough (we are ignoring transaction costs). To
avoid arbitrage, the discount rate therefore must be the risk-free
one. The portfolio $\underline{P}$ in this sense is perfectly
hedged over the short-term and we should set
$r_{\underline{P}}(v)=0$ in Eqs.\no{diffBS} and\no{tilder}.

\subsection{A Comment on Stochastic Volatility} The parabolic
partial differential equation\no{diffBS} was derived without
specifying a stochastic process for the asset's price. Specific
properties of the short-term returns enter the solution
to\equ{diffBS} only through the boundary conditions. Integration
of\equ{diffBS} for a European-style option requires knowledge of
the payoff of the option and of the overall variance $v_f$ of the
asset's returns at exercise. The option payoff is readily
expressed in terms of the value of a risk-free bond that matures
when the option expires. For European-style options the only
uncertainty thus is in the final variance $v_f$ of the asset's
returns.

By construction, the price of a European-style call option {\it
does not depend on the volatility path}. Paths with the same
overall variance $v_f$ of the asset's returns when the option
expires give the {\it same} fair option price.

\equ{diffBS} implies that the fair value of an European-style
option {\it for a given final variance} of\equ{condexp} is the
risk-neutral conditional expectation of the option payoff with the
pdf\footnote{With respect to the variance at expiration $v_f$, the
distribution\no{pBS} solves the "backward" evolution equation that
corresponds to\no{diffBS}. },
\be{pBS}
p_{BS}(y|x,v_f)=(2\pi\sqrt{v_f})^{-1}\exp[-(y-x+v_f/2)^2/(2v_f)]\
.
\ee

Denoting the risk-neutral marginal probability distribution for
the overall variance of the returns by\footnote{The ellipses again
represent any other pre-visible quantities (such as the current
spot price $x$).} $q(v_f|T,\dots)$, the risk-neutral probability
measure $Q$ for European-style options on the asset is given by
the pdf
\be{pdfQ}
p_Q(y|x,T,\dots)=\int_0^\infty dv_f q(v_f|T,\dots)
p_{BS}(y|x,v_f)\ .
\ee
The pdf $q(v_f|T,\dots)$ is the only ingredient that specifically
depends on market expectations. It therefore is not uniquely
specified by the process for the asset.  Since we do not have
options on a particular asset to every strike $K$, the market is
not complete. The distribution $q(v_f|T,\dots)$ thus cannot be
uniquely inferred from the observed option prices. One may,
however, hope to obtain a reasonable estimate by using a trial
distribution for $q(v_f|T,\dots)$ whose mean and variance are
calibrated to reproduce a few observed option prices. Note
that\equ{pdfQ} represents the risk-free measure $Q$ as a
positively weighted superposition of Gaussian distributions with
mean, $x-v_f/2$ and variance $v_f$. The mean and variance of each
Gaussian are strictly correlated. In the next section we will
argue that this unfortunately does not appear to be very realistic
and\equ{pdfQ} probably is not sufficiently general to reproduce
the observed smile.

\section{The Valuation of European Calls in Realistic
Markets} In deriving\equ{diffBS} we tacitly assumed that the
expectation in\equ{pricing} is meaningful. Since the fair value of
a call that is deep in-the-money approaches $S-K$, we see that for
$\Delta\neq 1$, the fair value of the portfolios we have been
considering essentially becomes proportional to the price of the
underlying $S=S_B e^y$ for large values of $y$. The expectation
in\equ{pricing} for such portfolios is finite only if the price of
the underlying has finite expectation,
\be{finexp}
{\bf E}^{p_h}[S/S_B]=\int_{-\infty}^\infty dy\, e^y p_h(y|x,v)
<\infty \ .
\ee
Together with the result of Appendix~A that the contribution from
higher moments becomes negligible in the limit $h\rightarrow 0_+$,
we thus find that\equ{diffBS} (with $r_{\underline{P}}(v)=0$)
holds for options on class~I assets only.

The historical distributions for equities\cite{St99},
indices\cite{Go99} and commodities\cite{Ma02} do not belong to
this class. Empirically the probability densities for short-term
returns have tails that fall off as a power in $x$ only. For time
intervals between 5~minutes and three weeks the
observed\cite{St99} pdfs of the returns on equities are all
shape-similar and well reproduced\cite{Sc02b} by a t-distribution
for $3$ degrees of freedom with mean $\bar y(h;v)=x+\m(h;v)$ and
variance $h$,
\be{t-dist}
p^{\rm emp.}_h(y|x,\m)\sim\frac{2 h^{3/2}}{\pi((y-\bar y)^2+h)^2}\
.
\ee

The integral of\equ{finexp} diverges in this case and the
valuation of the previous portfolios is all but meaningless: being
long a call apparently becomes a very attractive position --
unfortunately, the risk associated with this position is not
calculable. If the probability for exceptionally large
fluctuations is sufficiently great, the expected future value of
some portfolios no longer is determined by small fluctuations
about $y\sim x$, even as $h\rightarrow 0_+$. The short-term
expected value of the portfolios we have been considering in this
case mainly comes from the exceptionally large fluctuations, even
though these are not the most frequent. Truncating the
expansion\no{expansion} about $y=x$ in this case gives an
inaccurate representation of the portfolio's variation in price
and the derivation of\equ{diffBS} is no longer valid.

The damage can be contained by considering only portfolios that
are immune to large variations in the price of the underlying. It
is sufficient to restrict to portfolios whose value is uniformly
bounded by a finite constant $V_{\rm max}(P)$,
\be{bounds}
|V_P(x,v)|< V_{\rm max}(P), \qquad\forall x, v<v_f \ .
\ee
Examples of simple portfolios that are bounded in this manner are
a vanilla put or a covered vanilla call. For class~II assets one
can select those portfolios from the above set that are bounded
and without appreciable risk for sufficiently short periods of
time\footnote{This procedure is not possible for Paretian return
distributions with a divergent variance\cite{McC85}.}. Since the
set of portfolios that satisfy\no{bounds} is smaller than the set
of admissible portfolios in the case of class~I processes, it is
not surprising that the valuation of options on class~II assets is
less constrained.

The simplest bounded dynamic portfolio that is without appreciable
risk contains two European-style options on the asset that differ
in strike or time to expiration. We here discuss the case of a
portfolio of two covered calls, $\cc_1$ and $\cc_2$ with the same
expiration date but strikes $k_1$ and $k_2$ respectively.

The portfolio's fair value in bonds when the variance of the
return distribution is $v$ and the asset's log-price is $y$ can be
written,
\be{port2}
V_P(y,v)=\Delta_1\, \cc_1(y,v)+\Delta_2\, \cc_2(y,v)\ ,
\ee
where the fair price of a covered call is,
\be{pcov}
\cc_i(y,h)=c_i(y,h)-e^y,\qquad i=1,2
\ee
The weights $\Delta_1$ and $\Delta_2$ of the two covered calls are
chosen so that the portfolio's price does not change appreciably
for small variations of the asset's price about its current
log-price $x$,
\be{riskfree2}
\frac{\partial}{\partial y} V_{\underline{P}}(y,v)\Big|_{y=x}=0\ .
\ee
The weights,
\be{weights2}
\underline{\,\Delta}\,_1= \cc_2^{\,\prime}(x,v)\ ;\qquad
\underline{\,\Delta}\,_2=- \cc_1^{\,\prime}(x,v)\ ,
\ee
give one possible solution to\equ{riskfree2}. When the variance
increases by $h$, the portfolio $\underline{P}$ with
weights\no{weights2} assumes the value,
\ba{riskfreeport2}
V_{\underline{P}}(y,v+h)&=&\cc_1(y,v+h)\cc_2^{\,\prime}(x,v)-
\cc_2(y,v+h)\cc_1^{\,\prime}(x,v)\nonumber\\
&=&\left|
\begin{array}{lr}
\cc_1(y,v+h) &  \cc_1^{\,\prime}(x,v)\\
\cc_2(y,v+h) &  \cc_2^{\,\prime}(x,v)
\end{array}\right|\ ,
\ea
for a return on the asset of $y-x$.  About $y=x$ and $h=0$,
$V_{\underline{P}}(y,v+h)$ has the expansion,
\ba{expandport2}
V_{\underline{P}}(y,v+h)&=&V_{\underline{P}}(x,v)+h \left|
\begin{array}{lr}
\dot{\cc}_1(x,v) &  \cc_1^{\,\prime}(x,v)\\
\dot{\cc}_2(x,v) &  \cc_2^{\,\prime}(x,v)
\end{array}\right|+\nonumber\\
&&+\frac{(x-y)^2}{2}\left|
\begin{array}{lr}
\cc_1^{\,\prime\prime}(x,v) & \cc_1^{\,\prime}(x,v) \\
\cc_2^{\,\prime\prime}(x,v) & \cc_2^{\,\prime}(x,v)
\end{array}\right|+O(h^2,(x-y)h,(x-y)^3)\ .\nonumber\\
&&
\ea
Using that the value of the portfolio $\underline{P}$ is bounded,
its expected future price with the pdf\no{t-dist} and sufficiently
short time intervals is,
\ba{estimate2}
&&\int_{-\infty}^\infty \hskip-1em dy\, p^{\rm
emp.}_h(y|x,\m)V_{\underline{P}}(y,v+h)=\int_{\bar y-1}^{\bar y
+1}\hskip-1.5em dy\,p^{\rm emp.}_h(y|x,\m)V_{\underline{P}}(y,v+h)
+
O(h^{3/2})\nonumber\\
&&\qquad =V_{\underline{P}}(x,v)+h \left|
\begin{array}{ccc}
-1 & 0 & 1\\
\dot{\cc}_1(x,v) &  \cc_1^{\,\prime}(x,v) & \half \cc_1^{\,\prime\prime}(x,v)\\
\dot{\cc}_2(x,v) &  \cc_2^{\,\prime}(x,v) & \half
\cc_2^{\,\prime\prime}(x,v)
\end{array}\right|+O(h^{3/2}\ln(h),\m^2)\ ,\nonumber\\
&&
\ea
where we have assumed that the expected short-term return on the
asset satisfies\equ{meangrowth}. The determinant of the $3\times
3$ matrix is the result of combining the expectations of the two
determinants in\equ{expandport2}. Because the portfolio value is
immunized against large price fluctuations, the truncation of the
transition probability in\equ{estimate2} induces an error of order
$h^{3/2}$ only (see Appendix~A for details). For class~II
short-term returns, the valuation of a bounded Delta-hedged
portfolio thus effectively is reduced to the class~I case. One
similarly can show that the variance of $V_{\underline{P}}$ is of
order $h^{3/2}$ and that $\underline{P}$ therefore is without
appreciable risk.

In the limit $h\rightarrow 0_+$, the fair values of any two
covered European-style calls on a class~II asset thus satisfy,
\be{det0}
\left|
\begin{array}{ccc}
-1 & 0 & 1\\
\dot{\cc}_1(x,v) &  \cc_1^{\,\prime}(x,v) & \half \cc_1^{\,\prime\prime}(x,v)\\
\dot{\cc}_2(x,v) &  \cc_2^{\,\prime}(x,v) & \half
\cc_2^{\,\prime\prime}(x,v)
\end{array}\right|=0\ .
\ee
This is one partial differential equation for two unknown
functions. However, since the determinant vanishes only when the
corresponding system of linear equations is dependent, we can
disentangle\equ{det0} into two linear partial differential
equations for each covered call separately -- at the cost of
introducing a function $\alpha(x,v)$. Excluding the possibility
that the value of a covered call does not depend on the asset's
price, \equ{det0} is equivalent to the set of linear equations,
\be{lineq}
\left(
\begin{array}{ccc}
-1 & 0 & 1\\
\dot{\cc}_1(x,v) &  \cc_1^{\,\prime}(x,v) & \half \cc_1^{\,\prime\prime}(x,v)\\
\dot{\cc}_2(x,v) &  \cc_2^{\,\prime}(x,v) & \half
\cc_2^{\,\prime\prime}(x,v)
\end{array}\right)\left(
\begin{array}{c}
1\\ \alpha(x,v)-\half\\ 1
\end{array}\right)=\left(
\begin{array}{c}
0\\
0\\
0
\end{array}\right)\ .
\ee
The first of these equations is true for any $\alpha(x,v)$. The
latter two imply that call options on a class~II asset satisfy the
partial differential equation,
\be{pde}
\dot c(x,v)+(\alpha(x,v)-\half) c^\prime(x,v)+\half
c^{\prime\prime}(x,v)=\alpha(x,v) e^x\ .
\ee
At any given moment, $\alpha(x,v)$ can be expressed in terms of
the "Greeks" for any European-style call on the underlying,
\be{defa}
\alpha(x,v)=-\frac{\dot{\cc}(x,v)-\half\cc^{\,\prime}(x,v)+\half
\cc^{\,\prime\prime}(x,v)}{\cc^{\,\prime}(x,v)}\ ,
\ee
and in particular does not depend on the strike of the option.

Our considerations of course also apply to transition
probabilities that fall off more rapidly than\no{t-dist}. One
indeed recovers the partial differential equation\no{diffBS} of
Black and Scholes as the special case
\be{aBS}
\alpha(x,v)=0\ .
\ee
As noted before, since a call with strike $K=0$ is worth the stock
at exercise, $c_{-\infty}(x,v)=e^x$ must be a special solution
to\equ{pde} that does not depend on $\alpha(x,v)$. The
inhomogeneous term in\equ{pde} for the valuation of call options
therefore is a matter of consistency. Put-call parity implies that
a European-style put with the same strike and expiration date as a
call satisfies the homogeneous partial differential equation {\it
with the same} $\alpha(x,v)$. By repeating the arguments for two
covered calls with different expiration dates, one concludes that
$\alpha(x,v)$ also does not depend on the expiration date of an
option. $\alpha(x,v)$ in this sense is an universal function that
does not depend on specific properties of European-style options.

The function $\alpha(x,v)\neq 0$, can be viewed as a risk-premium
on a covered call (respectively a put). The reason for such a
premium is evident from the derivation: it represents the cost of
insuring a simple Delta-hedged portfolio with just one option
against large fluctuations in the price of the underlying. Note
that $\alpha(x,v)$ enters the evolution equation for options with
bounded payoffs as an effective cost-of-carry for the underlying
asset in the Black-Scholes world would.

This interpretation of $\alpha$ becomes evident if we consider the
stochastic process whose generator $\hat A$ is the evolution
operator in\equ{pde},
\be{generator}
\hat A(v)\phi(w,v) =\left\{\half \frac{\partial^2}{\partial w^2}
+(\alpha(w,v)-\half)\frac{\partial}{\partial w}\right\}\phi(w,v)\
.
\ee
The corresponding stochastic process is\cite{Ok00},
\be{stochproc}
dw=(\alpha(w,v)-\half)dv + dB_v
\ee
where $B_v$ denotes Brownian motion with zero mean and variance
$dv$. The measure $Q$ of\equ{fta} that corresponds
to\equ{stochproc} is unique\cite{SV79} as long as the drift
$\alpha(w,v)$ is finite for all $w,v$ and does not increase
faster\footnote{The interpretation of $\alpha$ as an effective
cost of carry makes this mathematical statement rather obvious:
nobody will hold an asset whose cost of carry grows faster than
its return.} than $|w|$ for $|w|\sim\infty$. Assuming this to be
the case, the fair price of a European-style option on a class~II
asset is uniquely specified by $\alpha(w,v)$ and the marginal
risk-free stopping distribution $q(v_f|T,\dots)$.

[The $\half$ in the drift-term of\equ{stochproc} does not appear
in the corresponding stochastic process for $n(v):=e^{w(v)}$,
which follows geometric Brownian motion\footnote{The mean drift
$\alpha(n,v)$ in\equ{nproc} should not be confused with the mean
return of the asset. The two are not even related: the drift
$\alpha(n,v)$ is due to large fluctuations in the price of the
underlying, not due to its mean return. The stochastic
process\equ{stochproc} is {\it not} the one followed by the
log-price $x(v)$ of the asset.}
\be{nproc}
\frac{d n}{n}=\alpha(n,v) dv + dB_v\ ,
\ee
with mean instantaneous drift $\alpha(w=\ln(n),v)$.]

A constant effective risk premium on options was recently
interpreted by Derman\cite{De02} as due to a stock's intrinsic
time-scale generated by short-term speculators. Although our
argument apparently is somewhat different, the rather similar
effect described here may have a common origin: large exceptional
fluctuations in the short-term returns of the underlying perhaps
can be traced to speculation.  The asymptotic power law fall-off
of the return distribution\equ{t-dist} has indeed
recently\cite{Ga02} been linked to the speculative actions of
large investors such as mutual funds.

It is difficult to compare a risk due to exceptionally large
fluctuations to any risk arising from "normal" fluctuations
described by the variance of a distribution.  How this exceptional
risk is valued furthermore depends on the perception of investors.
The function $\alpha(x,v)$ thus probably is specified only by the
observed option prices themselves. In the absence of options to
every strike and exercise date, the problem of calibrating
$\alpha(x,v)$ to the observed market prices is not complete.
Additional assumptions are required -- for instance that the
relative entropy to the Black-Scholes model is minimal\cite{Av97}.

To better visualize the effect a non-vanishing drift has on option
prices, let us consider constant $\alpha>0$. One can explicitly
solve\equ{pde} in this case and obtains that the overall
Black-Scholes variance $v_{BS}(\tilde k,\alpha;v_f)$ at expiration
implied by a European call is implicitly given by the relation,
\be{impl}
\ln(v_{BS}/v_f)+\frac{(\tilde k+v_{BS}/2)^2}{v_{BS}}=\frac{(\tilde
k+v_f (1/2-\alpha))^2}{v_f}\ .
\ee
Here $v_f$ is the total variance of the asset's returns at the
time of exercise of the option and $\tilde k=k-x=\ln(K S_B/N_B
S)=\ln\widetilde K$ is its discounted strike in terms of the spot
price of the asset.\\*

~\hskip3em \epsfig{figure=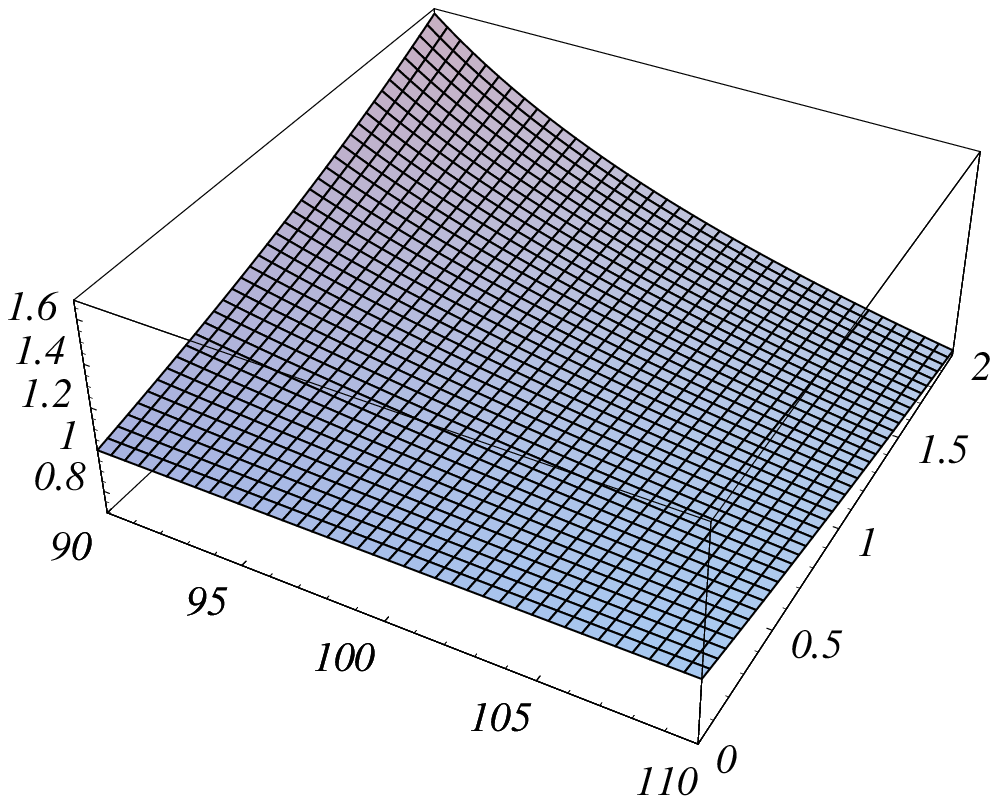,width=3in}\\*

\vskip-1.7in $v_{BS}(\%)$ \vskip0.3in\hskip3.3in
$\alpha$\vskip0.5in\hskip1in $\widetilde K(\%{\rm
spot})$\vskip0.3in \noindent{\small\baselineskip=7pt Fig.1: The
implied overall variance $v_{BS}(\widetilde K,\alpha;v_f)$ for
$v_f=1\%$. The surface is for European vanilla calls. The
drift-function $\alpha(x,v)$ in\equ{pde} is taken to be constant
and the discounted call strike $\widetilde K$ is in percent of the
spot price for the underlying asset.} \vskip10pt

A typical implied total variance surface for constant $\alpha\geq
0$ is shown in Fig.~1. For $\alpha=0$\equ{pde} reduces to the
Black-Scholes equation and $v_{BS}(\tilde k,\alpha=0;v_f)=v_f$. In
the Black-Scholes world with $\alpha=0$ there is no
volatility-skew if one conditions on the total variance $v_f$ of
the returns on expiry of the European-style options. As $\alpha$
increases the variance implied by deep in-the-money calls becomes
progressively greater compared to the one implied by
out-of-the-money calls. This behavior can be understood by
interpreting $\alpha$ as an effective cost-of-carry for the
underlying asset in the Black-Scholes world -- this cost evidently
is avoided by holding a deep in-the-money call instead of the
asset itself and is incurred when holding a deep in-the-money put.
The same reasoning also implies that for $\alpha>0$ in-the-money
calls should become more undervalued by the Black-Scholes model as
the time to expiration (and thus $v_f$) increases.   The simple
model with constant $\alpha$ thus tends to qualitatively reproduce
the volatility smile and term structure that is generically
observed for equities. The fact that the total variance $v_f$ of
the returns at the expiry of the option is itself uncertain has to
be taken into account in a realistic valuation. It does not
qualitatively change the picture if the risk-neutral uncertainty
of $v_f$ is not too great. A more richly structured variance
surface can be modelled by non-constant $\alpha(x,v)$.

\section{A Summary and Discussion of the Results}
We separated the problem of valuing European-style options from
that of constructing a risk-free portfolio by conditioning on the
overall variance of the asset's returns when the option expires.
For hedging purposes, assets with sub-exponential short-term
return distributions can be divided into those with finite and
infinite variance (classes~II and~III respectively). Most
financially interesting assets historically\cite{St99,Go99,Ma02}
belong to class~II. Fully dynamic portfolios that are risk-free
can be constructed for any asset with short-term returns of finite
variance and in particular for class~II assets. However, in
contrast to the Black-Scholes world, a risk-free portfolio in this
case has to be statically immunized against exceptionally large
fluctuations of the asset's returns. In effect this implies that
the portfolio's variation in value must be bounded. It therefore
contains at least two {\it different} European-style options on
the asset.

We use the variance overall $v$ of the returns on the underlying
instead of a "trading time". The final overall variance $v_f$ when
the European-style option expires is interpreted as a stochastic
"stopping time" for the risk-neutral diffusion. The diffusion for
the fair price of an option on a class~II asset was found to be
characterized by a drift $\alpha(x,v)$. Constant $\alpha>0$ for
given $v_f$ qualitatively reproduces the volatility smile and
term-structure often observed in equity markets. For short-term
return distributions that fall off exponentially or faster,
$\alpha=0$ and the diffusion reduces to the one of Black-Scholes,
albeit evolving in $v$ instead of in a trading time.

One might object to considering sub-exponential return
distributions for the underlying, since the price variations of an
asset, although perhaps large, can be thought of as restricted to
a finite range in the finite lifespan of the option. There are at
least two objections to this argument. Firstly, the value of
equities does sometimes change considerably in a short time and
fluctuations may exceed several standard deviations when a company
is forced into bankruptcy or announces new patents or
acquisitions. These scenarios are not so rare that they can be
disregarded in the valuation of options (except perhaps by
insiders). One realistically therefore may wish to immunize a
portfolio against large price-changes of the underlying. It also
is operationally and financially quite impossible to balance a
portfolio arbitrarily often. If the tails of the return
distribution are sufficiently fat, higher moments can become
relevant in the evaluation of\equ{pricing} when the change in
variance, $h$, between updates is finite. If higher moments of the
distribution of returns are sufficiently large (not necessarily
infinite) it again is advisable to immunize the portfolio against
large price fluctuations of the underlying asset, that is restrict
the portfolio's variation in value under large variations of the
asset's price. The same strategy was found to be useful in pricing
options on class~II assets: the portfolio in this case should be
statically immunized against exceptionally large fluctuations of
the underlying and dynamically hedged to make it insensitive to
normal ones as well.

Thus, although it may in reality not be possible to distinguish
sharply between assets of class~I and~II, the strategy employed
here for class~II assets is the more realistic one. The processes
for the two kinds of assets evidently can be continuously deformed
into each other and it is gratifying that the evolution equations
satisfied by the corresponding option prices also can be
continuously deformed from $\alpha(x,v)\neq 0$ (class~II) to
$\alpha(x,v)=0$ (class~I).

The returns on financial assets in realistic markets fortunately
have finite variance and the Paretian (class~III) scenario
therefore is quite academic. The risk-free measure for the
valuation of European-style options conditional on the final
variance of the asset's returns is unique for class~I assets. The
volatility smile and term structure in this case are entirely due
to the risk-free distribution of the final (stopping) variance. An
a priori unspecified drift function $\alpha(x,v)$ complicates the
valuation of European-style options on class~II assets.
By\equ{defa} $\alpha(x,v)$ is given by the "Greeks" of an option
and effectively measures the amount by which option prices at any
moment violate the Black-Scholes pde. We have interpreted
$\alpha(x,v)$ as the investor's compensation for the residual risk
of a single-option Delta-hedged portfolio due to exceptionally
large fluctuations in the asset's returns. As such, this drift is
not explicitly related to the process for the underlying. However,
quite interestingly, the effect of this drift on option valuation
is equivalent to that of an effective cost-of-carry for the
underlying asset in the Black-Scholes world.

{\bf Acknowledgements:} I am indebted to M.~Avellaneda for the
remark that the payoff of option portfolios in practice often is
"landscaped" to reduce the risk from large market movements and to
P.~Friz for a lengthy discussion. The continued encouragement and
support by L.~Spruch is greatly appreciated.

\appendix

\section{Estimates of the Remainders}
To justify the estimates in the text we here show that the
remainders in the expansions\no{expriskless} and\no{expandport2}
indeed are negligible as $h\rightarrow 0_+$.

We first consider a pdf $p_h(x)$ of class~I with variance $h$ and
vanishing mean. The argument of $p_h$ can always be shifted to
obtain a distribution with non-vanishing mean. Being in class~I
implies that,
\be{vevexp}
\int_{-\infty}^\infty dy\, p_h(y) e^{\lambda y}<\int_0^\infty dy
p_h(y) e^{y}+ \int_{-\infty}^0 dy p_h(y)<\infty\ ,\qquad \forall
\lambda<1\ ,
\ee
that is, the moment generating function is analytic about
$\lambda=0$ and all moments of $p_h(y)$, in particular, are
finite.

We are interested in integrals of the form
\be{expf}
\int_{-\infty}^\infty dy\, p_h(y)\, f(y)\ ,
\ee
for functions $f(y)$ that are analytic (almost) everywhere. The
remainder $R_N(y)$ in the McLaurin series
\be{Laurinf}
f(y)=\sum_{n=0}^N \frac{y^n}{n{\bf !}} f^{(n)}(0) + R_N(y)\ ,
\ee
thus vanishes as $N\rightarrow \infty$ for (almost) all $y$.  An
expression for the remainder is,
\be{remain}
R_N(y)=\frac{y^{N+1}}{(N+1){\bf !}}f^{(N+1)}(y\xi)
\ee
with $0<\xi<1$. Changing the scale of the integration variable
$y\rightarrow y\sqrt{h}$ the expectation of $R_N$ for small $h$
is,
\ba{vevN}
{\bf E}^{p_h}[R_N]&:=&\int_{-\infty}^\infty dy\, p_h(y) R_N(y)\cr
&~=&\frac{h^{(N+1)/2}}{(N+1){\bf !}}\int_{-\infty}^\infty dy\,
y^{(N+1)} \{\sqrt{h} p_h(y\sqrt{h})\} f^{(N+1)}(y\xi \sqrt{h})\ .
\ea
The pdf $\sqrt{h} p_h(y \sqrt{h})$ has unit variance and the
higher moments of the limiting pdf,
\be{lim}
p_1(y):=\lim_{h\rightarrow 0_+} \sqrt{h} p_h(y \sqrt{h})\ ,
\ee
are finite. Using that $f^{(N+1)}(x)$ is analytic about $x=0$, the
integral in\equ{vevN} has a finite limit for $h\rightarrow 0_+$
and one concludes that,
\be{vevNconclude}
{\bf E}^{p_h}[R_N]= O(h^{(N+1)/2})\ .
\ee
This estimate continues to hold when the function $f(x)$ is
expanded about a point $x=\m(h)$ that satisfies\equ{meangrowth}.
For analytic portfolio values, the estimates in Eqs.\no{BSvar} and
\no{expV2} thus are justified and the neglected terms are of
higher order in $h$.

The valuation of bounded portfolios with pdf's of class~II such
as\equ{t-dist} can be reduced to the previous case if the error
from truncating the pdf becomes negligible for $h\rightarrow 0_+$.
To see this, consider a pdf with zero mean and variance $h$ that
for sufficiently small $h$ is bounded by,
\be{maj}
p_h(y)\leq \frac{D}{\sqrt{h}}\left(\frac{h}{y^2}\right)^\nu \ \ \
{\rm for}\ |y|>1, h<h_0 \ ,
\ee
where $D>0$ and $\nu$ are constants that do not depend on $h$.
Note that if the variance of $p_h(y)$ is finite, \equ{maj} holds
for some $\nu>3/2$. The contribution of the tails of the
distribution  to the expectation of a bounded function $|f(y)|\leq
f_{\rm max}$ in this case is,
\be{tails}
\int_{|y|>1} dy\, p_h(y) f(y)\leq f_{\rm max}\int_{|y|>1} dy
p_h(y)\leq 2 D f_{\rm max} h^{\nu-1/2}\ .
\ee
The tails of any distribution in class~II (with $\nu>3/2$)
therefore give a sub-leading contribution to the expectation of a
bounded function and can be cut off. For bounded functions, a pdf
of class~II effectively can be replaced by one that vanishes for
$|y|>1$. This pdf of class~II with truncated tails is a pdf of
class~I up to a normalization factor. Using\equ{tails} with
$f(y)=1$, the normalization correction is of order $h^{\nu-1/2}$
and is itself sub-leading. The estimate of the expectation of
bounded functions for pdf's of class~II thus is reduced to the
previous case of class~I distributions. [Note that $\nu=2$ for the
realistic pdf of\equ{t-dist} -- the error induced by cutting off
the tails in this case is of the same order as that due to
neglecting the remainder in the expansion of the portfolio's
value.] For short-term returns with a finite variance, the
estimate of the order of the corrections in\equ{estimate2} thus is
justified for portfolios with bounded values and a distribution of
the returns on the asset that falls off like\equ{t-dist}.

\end{document}